

\magnification=1200\baselineskip=18pt
\def \r{{\bf r}}
\def \d{{\bf d}}
\def \v{{\bf v}}

\bigskip \bigskip \centerline {\bf From Feynman's Wave Function to the
Effective Theory of Vortex Dynamics} \bigskip
\centerline {Q. Niu} \centerline {Department of
Physics, University of Texas, Austin, TX 78712} \bigskip
\centerline { P. Ao, and D. J.
Thouless} \centerline {Department of Physics, University of Washington,
Seattle,
WA 98195} \bigskip \bigskip

We calculate the overlap between two many-body wave functions for a superfluid
film containing a vortex at shifted positions.  Comparing the results to
phenomenological theories, which treat vortices as point particles, we find
that
the results are consistent if the point-particle vortices are considered as
under the action of the Magnus force and in weak interaction with sound waves
of the superfluid.   We are then able to resolve the disagreement concerning
the
effective mass of vortices, showing it is finite.

\bigskip \bigskip \bigskip \bigskip

 \vfil \eject

Vortices play an important role in the understanding of both static and
dynamical properties of a superfluid[1]. They determine the
Kosterlitz-Thouless phase transition[2], and provide a mechanism for the mutual
friction between the superfluid and the normal fluid[3].
Due to advances in experimental techniques, there are  many
studies of problems related to  vortex dynamics, such as
the quantum nucleation of vortex rings induced by moving ions[4]
and quantum phase slippage near a submicron orifice[5].
In two dimensions, the theoretical framework for understanding these dynamical
phenomena is based on an effective point-particle formulation of vortex
dynamics, and has been very successful[1].
Naturally, physical quantities in the
phenomenological theory, such as the vortex mass, the Magnus force,
and the friction should be derived from a microscopic theory.
However, the current understanding of these quantities is
in a confused state: there is no clear calculation of the coupling of the
vortex to the low lying excitations responsible for the friction,
and the theoretical estimates of the vortex mass
range from zero[6], to finite[7], and to infinite[8].
There is also a suspicion
that an effective mass may not be meaningfully defined
for a vortex after all[9].

The purpose of the present paper is to present a conceptually
straightforward calculation to give clear constraints on these
quantities.  We invoke a microscopic
description of the vortex by writing a Feynman many-body wave
function for a superfluid film containing a vortex[10]. We calculate the
overlap
integral between such a state and that with the vortex shifted a distance away,
and find how it behaves as a function of the distance.
We also calculate the same quantity within the phenomenological
point-vortex theory.  Comparing the two, we conclude that the effective mass of
the vortex cannot be infinite, and that the coupling of a vortex with low lying
excitations must be sufficiently  weak. At the end of the paper, we will
discuss
the generality of our approach and its application to other systems.

Let us start with the phenomenological theory of vortex dynamics in a two
dimensional superfluid film.
A vortex is regarded as a point particle moving under the influence
of the Magnus force $h\rho_0 \hat z\times \v$, where $h$ is the Planck
constant, $\rho_0$ is the 2-d superfluid number density, $\hat z$ is the unit
vector normal to the film, and $\v$ is the velocity of the vortex.  Its
effective Hamiltonian may be written as
$$
  H_v = {1\over 2 m_v}[-i\hbar \nabla - q \; {\bf A}(\r )]^2, \eqno (1)
$$
where $m_v$ is an effective inertial mass of the vortex, the vector potential
$\bf A$ in the symmetric gauge for the Magnus force is $(-y, x)h\rho_0/2$, $\r$
is the vortex coordinate, and $q=\pm 1$ is the vorticity of the vortex.
Eq.(1) can be understood by drawing an analogy with the case of
a two dimensional electron moving under the influence of the Lorentz force by
a magnetic field, with $q$ interpreted as the vortex `charge'.[11]

This simple phenomenology is unfortunately not adequate if one wishes to
compare

with a more microscopic theory.  One must also include
interactions with low lying excitations such as various sound waves of the
superfluid, which may be realized by the following model interacting
Hamiltonian
$$
  H_i = q \sum_{\bf k} M(k)e^{i{\bf k}\cdot{\bf r}}
  (a_{\bf k}+ a_{-{\bf k}}^\dagger), \eqno(2)
$$
where ${\bf k}$ is the wave vector of a low lying excitation with
the corresponding creation
(annihilation) operator $ a^\dagger_{\bf k}$ ($ a_{\bf k} $). An index
labeling different kinds of excitations is omitted for notational
simplicity.  Coupling of this form conserves the total momentum of the system,
as is necessary for a translationally invariant system.
The Hamiltonian for the low lying excitations is
$$
  H_e = \sum_{\bf k} \hbar \omega_k   a_{{\bf k}}^\dagger a_{\bf k}
                  . \eqno(3)
$$
Therefore the total Hamiltonian of the system, a vortex and the low
lying excitations, is
$$
  H = H_v + H_i + H_e . \eqno(4)
$$
In the following, we will focus our attention on the overlap integral between
different vortex states, in which the vortex mass $m_v$, the Magnus force,
and the coupling between the vortex and the low lying excitations should be
involved.

In the absence of the coupling to the low lying excitations,
the overlap integral between two coherent states for a vortex centered at
${\bf r}'_{0}$ and ${\bf r}_0$
in the ground state can be calculated as[11]:
$$
  O({\bf r}_0, {\bf d}) = <\r_0'|\r_0 > = \exp\left[- {1\over
  4l_m^2}|\d|^2+{i\over 2\l_m^2}\hat z\cdot (\d\times\r_0) \right],\eqno(5)
$$
where $l_m=(2\pi\rho_0)^{-1/2}$ is the
mean spacing between the atoms in the superfluid, and
the vector ${\bf d} = {\bf r}_{0} - {\bf r}'_0$.
The coherent state has the form of eq.(6) below,
with $l$ and $l'$ replaced by $l_m$, and $|\psi_e>$ by the vacuum
of the low lying excitations.  The above overlap integral contains a phase
facto
   r,
derived from the Berry phase (or Aharonov-Bohm phase in this context) of the
coherent state.  It also contains  a gaussian decay factor, reflecting the
localization of the coherent state.   Both factors are characterized by the
same
length scale $\l_m$, and are independent of the vortex mass $m_v$.

In the presence of interactions with the low lying excitations,
the total overlap integral will change in two ways by the polaron effect[12]:
(1) The vortex can induce polarization of the excitations, and the
overlap between
the polarized excitations of one coherent state of the vortex and those of a
shifted coherent state can contribute to the reduction of the total overlap
integral.
(2) The polarized excitations tend to localize the vortex, squeezing the
coherent state to a smaller size than $l_m$.  These effects will clearly depend
on the interaction strength, and will also involve the vortex mass.
Now a coherent state of the vortex
centered at $\r_0$ may be approximated by the following variational wave
functio
   n
$$
  |\r_0>= {1\over \sqrt {2\pi l^2} }
  \exp \left[-{|\r-\r_0|^2\over 4l^2}+{i\hat z
  \cdot \r_0\times \r
  \over 2 l'^2}\right] \times |\psi_e> , \eqno (6)
$$
where $l$ and $l'$ are two variational parameters, and
$|\psi_e>$ is a wave function of the excitations only.
With the above ansatz, the total energy of the system is evaluated as
$$
  E = {\hbar^2\over 4m_v l_m^2}
  \left[ {l^2 \over l_m^2 } + {l_m^2 \over l^2 }
         + { \r_0^2 \over 2 } \left( {1 \over l_m } - { l_m \over l'^2 }
                     \right)^2 \right]
   + <\psi_e|(H_e+\bar H_i)|\psi_e>, \eqno(7)
$$
where
$$
  \bar H_i = q \sum_{\bf k} M(k) e^{- { k^2l^2 \over 2 } }
  e^{i{\bf k}\cdot{\bf r_0}} (a_{\bf k} + a_{-{\bf k}}^\dagger).\eqno(8)
$$
First, the energy is minimized by taking $|\psi_e>$ as the ground state of
$H_e+\bar H_i$, namely
$$
  |\psi_e> = \exp\left[ q \sum_{\bf k} {M(k) e^{-k^2l^2/2}\over
  \hbar\omega_k} e^{i{\bf k}\cdot{\bf r_0}}
  (a_{\bf k} - a_{-{\bf k}}^\dagger)\right] |0>,\eqno (9)
$$
where $|0>$ is the vacuum of the excitations.
The energy of the system then becomes
$$
  E = {\hbar^2\over 4m_v l_m^2}
  \left[ { l^2 \over l_m^2 } + { l_m^2\over l^2}
   + {\r_0^2\over 2} \left({1\over l_m}-{l_m\over l'^2}\right)^2\right] -
  \sum_{\bf k} {|M(k)|^2 e^{-k^2 l^2}\over \hbar \omega_k}.\eqno(10)
$$
Obviously $l'=l_m$ minimizes eq.(10).
The energy is further minimized with respect to $l$ if
$$
  l^{-4} = l_m^{-4}+ {4m_v \over \hbar }
  \int_0^\infty d\omega {J(\omega) \over \omega} e^{-k_{\omega}^2 l^2} ,
\eqno(11) $$
where the spectral function $J(\omega)$ is defined as
$$
  J(\omega) = \sum_{\bf k} {|M(k)|^2 k^2 \over \hbar^2}
                       \delta (\omega_k-\omega) . \eqno(12)
$$
Having the variational parameters $l$ and $l'$ determined,
the overlap integral $O(\r_0, \d)$ is then found as
$$
  O(\r_0, \d) = <\r'_0|\r_0> = \exp\left[- {1\over 4l_d^2} \d^2 + {1\over
       2\l_m^2}(i\hat z\cdot \r_0\times \d) \right] ,
      \eqno(13)
$$
for a sufficiently small distance $|\d |$.
Here the decay length $l_d$ in eq.(13) is
$$
  {1\over l_d^2} = {1\over 2 l^2}+ {l^2 \over 2l_m^4}
   +  \int_0^\infty d\omega {J(\omega) \over \omega ^2}
     e^{-k_\omega^2 l^2} . \eqno(14)
$$

The above results have several interesting features.  First, the length
in the Berry phase term is not renormalized by the interactions.
In fact, the same result is
reached even if we assume a general phase factor in the ansatz eq.(6).
The result has nicely
demonstrated the robustness of the  Berry phase term
against the details of the system.
Secondly, by eq.(11) the localization  length $l$ is smaller than $l_m$.
The effective mass $m_{v}$ enters in the equation,
because it determines the Landau level spacing, which in turn tells
how hard it is to mix with the higher Landau levels in order to shrink $l$.
Thirdly, the last term of the decay length of the overlap integral, eq.(14),
comes from the overlap of the polarized excitations.
Finally, we should point out that
when we consider the contribution from the fluctuating vector potential in
eq.(1) all these features remain unchanged.
Eqs.(11-14) are the results for the overlap integral
from the consideration of the effective theory.

Now we turn to a completely different way of obtaining the overlap integral,
a microscopic calculation based on Feynman's many-body wave function.
We will show that there is a complete correspondence between the two
approaches.  This will enable us to determine the vortex mass, the
Magnus force, and the coupling to the excitations.
If $\psi_0(\r_1...\r_N)$ is the ground state many-body wave function of
He II, the system  with a vortex
centered at position $\r_0$ may be described in a first approximation by[10]
$$
  |\psi(\r_0)> = \prod_{j=1}^N
  \exp[i\theta(\r_j-\r_0)+\alpha(\r_j-\r_0)]\psi_0(\r_1...\r_N), \eqno(15)
$$
where $\theta(\r)$ is the angle of $\r$, and $\alpha(\r)$ is a real
function of $|\r|$.
The most interesting feature of the wave function is that
it changes phase by $2\pi$
whenever an atom moves around the vortex center once.
In fact, it is by this feature that a vortex state should be defined;
the above wave function should be regarded as an approximate
description of the lowest energy state with this feature.
The phase factors in eq.(15) introduce a singularity
to the many-body wave function at the vortex center,
and this must be canceled by requiring $\exp[\alpha(\r)]$
to  vanish at the origin,
otherwise the cost in kinetic energy would be too high.  The particle
density in the state, eq.(15), therefore vanishes at $\r_0$.
At large distances, the depletion of
particle density due to the vortex vanishes like  $|\r-\r_0|^{-2}$,
and correspondingly $\alpha$ decays to zero like $|\r-\r_0|^{-1}$[13].

The full calculation of the overlap integral from the many-body wave function
is difficult, but we may expand $\ \ln O(\r_0,\d)$ in powers of $\d$ in the
small $|\d |$ limit.
In this limit the two leading terms only involve
one- and two-body density distributions in the state eq.(15) as will be shown
below.
Concrete results will then be obtained from a comparison with same leading
terms in eq.(13).
To facilitate the expansion we write
$$
  O(\r_0,\d) = < \psi(\r'_0)|\psi(\r_0) >
             = <\exp\left\{
    \sum_j[ig_1(\r_j-\r_0,\d)+g_2(\r_j-\r_0,\d)] \right\}>, \eqno(16)
$$
where $<>$ denotes average over the state of eq.(15), and we have used the
notation that $g_1(\r,\d)=[\theta(\r+\d/2)-\theta(\r-\d/2)]$,
and that $ g_2(\r,\d)=
\alpha(\r-\d/2)+\alpha(\r+\d/2)-2\alpha(\r).$  Up to second order in $\d$,
we may write $g_1(\r,\d) = \d\cdot \hat z \times (\r-\r_0) / |\r-\r_0|^2 $
and $g_2(\r,\d)={1\over 4}(\d\cdot\nabla)^2 \alpha(\r)$.
A straightforward cumulant  expansion of eq.(16) then yields, to
the same order in $\d$, that
$$\eqalign{
  \ln O(\r_0,\d)&=\int d^2\r\, \rho(\r)\,
  \left[i{\d\cdot \hat z \times (\r-\r_0)\over |\r-\r_0|^2 } \right]
  + \int d^2\r\, \rho(\r)\, \left[ {1\over 4} (\d\cdot\nabla)^2
  \alpha(\r-\r_0)\right]\cr
  &-{1\over 2}\int d^2\r\,\rho(\r)\,
  \left[{\d\cdot \hat z \times (\r-\r_0)\over |\r-\r_0|^2 }\right]^2
  \cr & -{1\over 2} \int\!\int d^2\r d^2\r' \,\rho(\r,\r')\,
  {\d\cdot \hat z \times (\r-\r_0)\over |\r-\r_0|^2 }\,
  {\d\cdot \hat z \times (\r'-\r_0)\over |\r'-\r_0|^2 }\cr
  & + {1\over 2} \left[\int d^2\r\,
  \rho(\r)\, {\d\cdot \hat z \times (\r-\r_0)\over |\r-\r_0|^2 }\right]^2,\cr}
  \eqno(17) $$
where $\rho(\r)\equiv<\sum_j \delta(\r-\r_j)>$ and $\rho(\r,\r')\equiv
<\sum_{i\ne j} \delta(\r-\r_i)\delta(\r'-\r_j)>$
are the one- and two-body density distributions in the state eq.(15).

The first order contribution to $\ln O(\r_0,\d)$ in eq.(17)
is purely imaginary, which can be evaluated
as $i\pi\rho_0\hat z\cdot (\d\times \r_0)$ if we replace $\rho(\r)$ by
$\rho_0$,
assuming that our system (including the  vortex center) is
confined  within a disc centered at the origin of $\r$. Here $\rho_0$ is the
2-d superfluid number density.
The correction due to $\rho(\r)-\rho_0\equiv\rho_1(\r)$ is zero
in the infinite size limit,
because of the rotational symmetry in $\rho_1(\r)$ about $\r_0$
and the fact that the density
depletion decays sufficiently fast at large distances from $\r_0$.
This first order term is the Berry phase associated with the Magnus force
discussed in Ref.[14].

The second order contribution to $\ln O(\r_0,\d)$ in eq.(17) is  purely real,
and it must also be
negative as required by the fact that $|O(\r_0,\d)|<1$ for nonzero $\d$.
We may therefore write
$$
  \ln O(\r_0,\d)=i\pi\rho_0\hat z\cdot
  (\d\times \r_0)- {\d^2 \over 4l_d^2} + {\rm higher\ order\ terms},\eqno(18)
$$
where we have put the second order term as independent of the direction of $\d$
because of the isotropy of the system about the vortex center.
The second order coefficient has been  parameterized by $l_d$,
which has the dimension of a length, and represents the same decay length as in
eq.(13).

We now examine closely the second order terms in eq.(17), and show that their
contribution to $l_d^{-2}$ is finite.
The term containing $\alpha(\r-\r_0)$ converges
because the double derivative of $\alpha$ decays as
an inverse cubic function at large distances from the vortex center
while $\rho(r)$ approaches a constant.
At short distances, $\alpha$ may diverge like a logarithm,
but $\rho(r)$ vanishes linearly,
causing no trouble to the convergence of the integral. Therefore we shall no
longer consider this term.
In the presence of particle correlation, the form of $\rho(\r,\r')$ is
unknown for the state containing a vortex,
except at large distances away from $\r_0$, where  it
reduces to $\rho_0(\r-\r')$, the distribution in the absence of the vortex.
We may, however, replace the distributions by their asymptotic forms
in eq.(17) in order to examine the long distance contributions to these terms,
because it is only from there that a divergence may ever be possible.
Then, the last
three terms of eq.(17) ({\it c.f.} eq.(18)) yield
$$
  {{\bf d}^2 \over 4 l_d^2} =  {1 \over 2 }
       \int {d^2{\bf k} \over (2\pi)^2} \rho_0 S_0({\bf k})
       |F({\bf k})|^2+..., \eqno(19)
$$ where `...' stands for the correction due to short distance
contributions, $S_0({\bf k})$ is the static structure factor in the absence of
t
   he
vortex, and $F({\bf k})=i2\pi e^{i{\bf k} \cdot \r_0} \hat z\cdot \d\times {\bf
k}/ k^2$ is the Fourier transform of $\d\cdot \hat z \times (\r-\r_0) /
|\r-\r_0|^2 $. It is known[13] that $S_0({\bf k})={\hbar k /  2Mc} $ for small
$
   k$,
where $M$ is the mass of a helium atom, and $c$ is the sound velocity.
The integral in eq.(19) therefore converges,
meaning that the second order expansion in eq.(17)
exists in realistic situations.

Before we proceed further we would like to comment on the validity of the above
discussions.  We have ignored multi-particle correlations  induced by
the vortex in the original wave function eq.(15) and in the evaluation of the
expression eq.(17).  We assume that the induced correlations decay sufficiently
fast away from the vortex center, such that they do not affect the convergence
properties at large distances.  The situation at short distances is very
complicated[10], and the short distance contribution can be quite substantial
to the reduction of the overlap function.  We expect, however, that the system
should behave smoothly at short distances, so that no divergence can be induced
from there.  Our later arguments will only be based on the conclusion drawn
above that the  decay length $l_d$ is finite.

The decay length $l_d$ strongly depends on the interaction
between the atoms in the superfluid.  As the interatomic interaction becomes
weak, the sound velocity decreases,
which makes $S_0({\bf k})$ large and therefore $l_d$ small from eq.(19).
In the extreme case of no interatomic interaction, $l_d$ becomes zero.
This is just what one should expect from a direct calculation of eq.(17)
in the free boson limit, in which case $\rho(\r,\r')=\rho(\r)\rho(\r')(N-1)/N$.

With the overlap integral evaluated both from the effective theory, eq.(13),
and Feynman's many-body wave function, eq.(18),
we now would like to see how the parameters of the effective theory
should be constrained.
Firstly, the parameter $\rho_0$ in the Magnus force of the effective theory is
the same as the 2-d superfluid density from the comparison of the results for
the Berry phase term of the overlap integral.  Second, in order to be
consistent
with the result of finite $l_d$ from Feynman's many-body wave function,
eq.(19),
the sum in eq.(14) has to be convergent, implying that the spectral function
$J(\omega)$ must vanish faster than $\omega$ at low frequencies and
$|M(k)|$ must be less singular than $k^{-1}$ at small $k$'s.
A comparison of eqs.(19) and (14) suggests that in the low frequency limit
$$
  J(\omega) =  {h \rho_0 \over 2M c^2 }   \omega^2 . \eqno(20)
$$
In the language of quantum theory of dissipation[15],
this kind of coupling is of the so-called superohmic type.
In a recent study of vortex tunneling in Ref.[16],
a general heat bath is considered.  It is found there that
a superohmic coupling to the heat bath has a negligible effect on the
tunneling process at low enough temperatures.

As for the mass of the vortex,
our result of finite decay length implies that
the mass of the vortex cannot be infinite,
otherwise the localization length
of the vortex would shrink to zero according to eq.(11) and the decay length
of the overlap function would become zero according to eq.(14).
Therefore, our result is consistent with that of Refs.[6,7], which suggest that
$m_v$ is zero or finite, and is in apparent disagreement with that of Ref.[8]

The vortex mass that we originally introduced in eq.(1) may have
already included the effect of renormalization
by the polarization of all but the low lying excitations of the superfluid.
There is still a
possibility that it may be renormalized  to infinity if the polarization
of the low lying excitations is included.
Indeed, if we neglect the Magnus force, a
straightforward perturbative calculation[12] shows that
the mass renormalization  becomes
logarithmically divergent if the coupling spectrum
$J(\omega)$ goes as $\omega^2$ at low frequencies.
This is essentially the result in Ref.[8].
The divergence of the mass renormalization becomes severer if
$J(\omega)$ would vanish slower than $\omega^2$.
However, this will not be the case for a dynamical process with a time scale,
as the following arguments show.


The situation in the presence of the Magnus force is quite different.
One can no longer set up a momentum eigenstate
and extract an effective mass of the vortex from the energy dependence
on the momentum.  A more natural approach is to relate the effective mass
to higher Landau levels of cyclotron frequency $\omega_c=\hbar\rho_0/m_v$.
Interaction with low
lying excitations may shift and broaden the higher Landau levels,
but these effects are not
divergent in a perturbative calculation using eq.(2)
if the coupling is superohmic.  Therefore,
if the higher Landau levels are well defined before turning on the coupling to
the low lying excitations, we can conclude that further inclusion of
such coupling has little
effects on the higher Landau levels and thus the effective mass of the vortex.
To observe a higher Landau level experimentally,
one may trap ions in vortices produced in a rotating film of superfluid,
and excite the vortices by electrical coupling to the ions.[17]

Finally, we would like to make some remarks about the generality of our
results.
As long as the Feynman many-body wave function description
of the vortex state is valid, everything else
just follows from standard many-body physics such as the form of $S_0({\bf k})$
at small $k$. As long as $S_0({\bf k})$
 vanishes with some positive power of $k$,
orthogonality catastrophe in the overlap integral will not occur. Therefore,
our
results may also be applicable to vortex structures in  superconducting films
and wire networks, Josephson junction arrays, and quantum spin systems.

\bigskip \noindent {\bf  ACKNOWLEDGMENT}   \bigskip

QN wishes to thank E. Fradkin, X.G. Wen, and D. H. Lee for helpful
discussions, and to thank M. Marder for a critical reading of the manuscript.
This work was supported in part by the Texas Advanced Research Program, by the
Robert A. Welch Foundation,  and by the National Science Foundation under Grant
No's. DMR 89-16052 and 92-20733. Two of us (QN
and DJT) are grateful for the hospitality of the Aspen Center for Physics.

\bigskip  \noindent {\bf REFERENCES}  \bigskip

\item{[1]} R.J. Donnelly, {\it Quantized Vortices in Helium II},
        (Cambridge, Cambridge, 1991)
\item{[2]} J.M. Kosterlitz and D.J. Thouless, J. Phys. {\bf C6}, 1181 (1973).
\item{[3]} W. F. Vinen, Proc. Roy. Soc. {\bf A240}, 114, (1957);
        A.L. Fetter, in {\it The Physics of Liquid
        and Solid Helium}, ed. by K.H. Bennemann and J.B. Ketterson
                                (Wiley, New York, 1976.)
\item{[4]} P.C. Hendry {\it et al.},  Phys. Rev. Lett. {\bf 60}, 604 (1988).
\item{[5]} J.C. Davis  {\it et al.},  Phys. Rev. Lett. {\bf 69}, 323 (1992);
           G.G. Ihas   {\it et al.}, {\it ibid,} { 327}.
\item{[6]} G. E. Volovik, JETP Lett. {\bf 15}, 81  (1972).
\item{[7]} C. M. Muirhead, W. F. Vinen, and R. J. Donnelly,
        Phil. Trans. R. Soc. Lond. {\bf A 311}, 433 (1984).
        Their opinion is based on the result of classical
        hydrodynamics that even a hollow vortex has an inertial mass of
        the fluid expelled by the vortex, see H. Lamb, {\it Hydrodynamics},
        pp76-80 (Dover, NY, 1945).
        Also, see, G. Baym and E. Chandler, J. Low Temp. Phys. {\bf 50}, 57
        (1983).
\item{[8]} J.M. Duan, Phys. Rev. {\bf B48}, 333 (1993);
        J.M. Duan and A. J. Leggett, Phys. Rev. Lett. {\bf 68}, 1216 (1992).
\item{[9]}  D.H. Lee, X.G. Wen, and S.M. Girvin, private communications.
\item{[10]}  R. P. Feynman, in {\it Progress in Low
        Temperature Physics I} , C.J. Gorter, ed.( North-Holland,
        Amsterdam, 1955);
        R. P. Feynman and M. Cohen, Phys. Rev. {\bf 102}, 1189 (1956);
        D.J. Thouless, Ann. Phys.(NY) {\bf 52}, 403 (1969).
\item{[11]} A. Perelomov, Theor. Math. Phys. {\bf 6}, 156 (1971).
\item{[12]}  G.D. Mahan, {\it Many-Partical Physics}, (Plenum, New York, 1981).
\item{[13]} P. Nozieres and D. Pines, {\it The Theory of Quantum Liquids},
        Vol.II, pp 68-71 (Addison-Wesley, NY, 1990).
\item{[14]} F.D.M. Haldane and Y.-S. Wu, Phys. Rev. Lett.
        {\bf 55}, 2887 (1985);
        D. Arovas, J.R. Schrieffer, and F.
        Wilczek, Phys. Rev. Lett. {\bf 53}, 722 (1984);
        P. Ao and D. J. Thouless, Phys. Rev. Lett. {\bf 70}, 2158 (1993).
\item{[15]} A.O. Caldeira and A.J. Leggett, Ann. Phys.(NY)
         {\bf 149}, 374 (1983);
         A.J. Leggett {\it et al.}, Rev. Mod. Phys. {\bf 59}, 1  (1987).
\item{[16]} P. Ao and D. J. Thouless, preprint, (1993).
\item{[17]} B. E. Springett, Phys. Rev. {\bf 155}, 139 (1967);
         W. P. Pratt, Jr., and W. Zimmermann, Jr.,
         {\it ibid}, {\bf 177}, 412  (1969).
\vfil\eject \end